# portable probe for photoacoustic imaging *in vivo*

Yongjian Zhao, Shaohui Yu, Luyao Zhu, Hengrong Lan, Jinwei Li, Jianfeng Li, and Fei Gao, *Member, IEEE*

*Abstract—* A low-cost adjustable illumination scheme for hand-held photoacoustic imaging probe is presented, manufactured and tested in this paper. Compared with traditional photoacoustic probe design, it has the following advantages: (1) Different excitation modes can be selected as needed. By tuning control parameters, it can achieve bright-field, dark-field, and hybrid field light illumination schemes. (2) The spot-adjustable unit (SAU) specifically designed for beam expansion, together with a water tank for transmitting ultrasonic waves, enable the device to break through the constraints of the transfer medium and is more widely used. The beam-expansion experiment is conducted to verify the function of SAU. After that, we built a PAT system based on our newly designed apparatus. Phantom and *in vivo* experimental results show different performance in different illumination schemes.

## I. INTRODUCTION

Photoacoustic imaging (PAI), also known as optoacoustic imaging (OAI), as an emerging technology in biomedical imaging, is developing dramatically fast in recent years [1] [2]. The primary imaging mechanism can be summarized as the following processes: The chromogenic group in tissue is illuminated by short-pulse laser, and the tissue that is sensitive to a specific wavelength of laser produces rapid thermal expansion and contraction, creating ultrasonic waves. The waves generated by inner tissue can be obtained by the ultrasonic transducer for image reconstruction. This hybrid imaging method combines the advantages of conventional optical with those of acoustic imaging [3] [4], i.e. the high contrast of optical imaging and the depth-resolved high resolution of ultrasound imaging at vast tissue depths[5] [6].

Limitations such as high cost, bulky size, and mode-fixed illumination of current PA imaging equipment have restricted further development in clinical application[7] [8]. For laser illumination, the schemes of PA imaging system can be divided into dark- and bright-field. The dark-field illumination is capable of minimizing surface interference signals and reducing their contributions to the background of deeper signals[9]. Different combinations of parameters produce different illumination schemes to achieve different illumination patterns. However, each type of illumination should still have its own advantages under certain circumstances. For PA image reconstruction, various illumination schemes directly affect the generation of PA signal and image quality[10] [11].

In this work, a new light-adjustable PA imaging system (LaPAI) was proposed, which can provide a versatile bright field, dark field illumination schemes, or even the combination of multiple field dynamicly, according to different application needs. We will firstly introduce the design of the spot-size adjustable Unit (SAU) with mathematical modeling and the structure of the PA probe. Then, we fabricated a vessel phantom by 3D printing to validate the imaging performance of the PA probe. Moreover, *in-vivo* experiments were executed for verification of the imaging capacity.

## II. SYSTEM DESIGN

### A. Probe design

To continuously adjust the spot size, we designed a beam expander based on the optical principle of the Galileo telescope, with a continuous zoom function. The unit consists of three lenses shown in Fig.1, where $L_1$ and $L_3$ are convex lenses, and $L_2$ is a concave lens. We can think of lenses $L_1$ and $L_2$ as combined lenses [12]. The combined focal length $f'_{comb}$ can be given by the following equation.

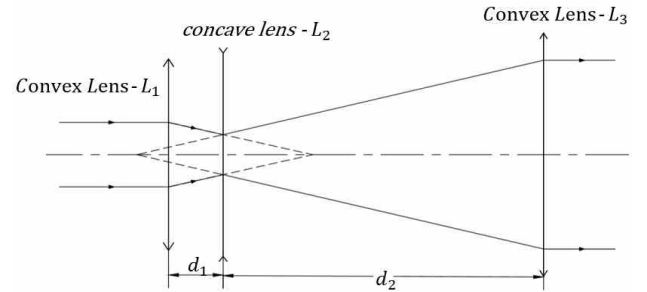

Fig.1 Lens design schematic in SAU.

$$f'_{comb} = \frac{f'_1 f'_2}{f'_1 + f'_2 - d_1} \quad (1)$$

Where $f'_1$ and $f'_2$ are the focal lengths of $L_1$ and $L_2$. The below equation gives the magnification of the spot diameter:

$$M = \frac{f'_3}{f'comb} \quad (2)$$

When $f'_1$ and $f'_2$ are determined, based on the object image exchange principle, the shortest focal length and the longest focal length of the compensation unit are in the same position during the entire focal length change, and the effect on the total focal length is the same, without affecting the multiple change ratio. We introduce $m_2$ to describe the combined focal length.

$$f'_{comb} = f'_1 m_2 \quad (3)$$

Generally, we follow the principle that each component does not collide during the motion. The parameter at the long focal length position is always used as the starting position for calculating the positive group compensation. Therefore, $m_{2l}$ and $d_{1l}$ can be used as initial data ($m_{2l}$ represents the zoom ratio corresponding to the variable length group $L_2$ of the system long focus, and $d_{1l}$ represents the spacing between the lens groups $L_1$ and $L_2$ at the long focus). For the positive compensation focus-adjustable system, the two parameters $m_2$ and $d_1$ determine the Gaussian solution range of the system, and the size of the variable-ratio $N$ only determines the length of the selected segment on the compensation curve and does not determine the range of the Gaussian solution [12].

Next, we determine the motion trajectory of the compensating lens $L_1$ and the variator lens $L_2$. During the whole motion, the movement amounts of $L_1$ and $L_2$ are $\Delta x_1$, $\Delta x_2$ respectively, and the object point is unchanged during the movement of $L_2$ whereas the image point moves. The movement of $L_1$ is utilized to compensate for the image plane. According to the zoom theory, we know that the algebraic sum of the image plane movement due to the movement of $L_1$ and $L_2$ should be 0.

$$d_\eta = (1 - m_1^2) m_2^2 d_{\Delta x_1} \quad (4)$$

$$d_\varepsilon = (1 - m_2^2) d_{\Delta x_2} \quad (5)$$

$$(1 - m_1^2) m_2^2 d_{\Delta x_1} + (1 - m_2^2) d_{\Delta x_2} = 0 \quad (6)$$

Where $d_\varepsilon$ is a thimbleful movement of the image point due to the slight movement of $L_2$. $d_\eta$ is the movement of image point because of the motion of $L_1$, which can compensate for the whole image plane. Transforming the above differential equation into an equation with $m$ as an independent variable gives:

$$\frac{1 - m_1^2}{m_1^2} f'_1 dm_1 + \frac{1 - m_2^2}{m_2^2} f'_2 dm_2 = 0 \quad (7)$$

$$d_{\Delta x_1} = \frac{f'_1}{m_1^2} dm_1 \quad (8)$$

$$d_{\Delta x_2} = f'_2 dm_2 \quad (9)$$

The zoom unit of the system consists of two motion groups, $L_1$ and $L_2$, and the object point of the zoom core is at infinity. During zooming, the start and endpoints of component $L_2$ satisfy the object image exchange principle, so the start and end positions of element $L_1$ are the same. Here, only the movement of $L_2$ acts as a zooming effect, which is a zooming group, and the range of $m_2$ can be determined by the magnification in long and short focus:

$$-\frac{1}{\sqrt{N}} \leq m_2 \leq -\sqrt{N} \quad (10)$$

Where $N$ is blanket magnification of system. So the movement curve of the multiple-adjustable group is expressed as:

$$\Delta x_2 = \int_{-\frac{1}{\sqrt{N}}}^{-\sqrt{N}} (1 - m_2^2) d_{\Delta x_2} = f'_2 m_{2l} (\frac{1}{\sqrt{N}} - \frac{1}{\sqrt{N'}}) = f'_2 \left( \frac{f' - f'_{min}}{\sqrt{f'_{max} f'_{min}}} \right) \quad (11)$$

Where $N = \frac{f'_{max}}{f'_{min}}$, $N' = \frac{f'_{max}}{f'}$, so the trajectory of the multiple-adjustable lens is straight line.

Next we need to determine the Gaussian solution of the compensation curve. For the total differential equation (11), we can determine that the general solution is:

$$U(m_1, m_2) = f'_1 \left( \frac{1}{m_1} + m_1 \right) + f'_2 \left( \frac{1}{m_2} + m_2 \right) = C \quad (12)$$

When $L_1$ gets to the long focal length position, the solution is given by following equation:

$$f'_1 \left( \frac{1}{m_{1l}} + m_{1l} \right) + f'_2 \left( \frac{1}{m_{2l}} + m_{2l} \right) = C \quad (13)$$

Where $m_{xl}$ is value of the long focal length position. Using equation (12) to subtract (13), we got the new equation consisting of Magnification $m_1$ which comes from compensation group.

$$m_1^2 - b_{m_1} + 1 = 0 \quad (14)$$

$$b = -\frac{f'_2}{f'_1} \left( \frac{1}{m_2} - \frac{1}{m_{2l}} + m_2 - m_{2l} \right) + \left( \frac{1}{m_{1l}} + m_{2l} \right) \quad (15)$$

so we can get the solution of $m_1$:

$$m_{1,1} = \frac{b + \sqrt{b^2 - 4}}{2}$$

$$m_{1,2} = \frac{b - \sqrt{b^2 - 4}}{2} \quad (16)$$

From the analysis results, we can see that the Gaussian solution of the qualified compensation curve is not unique, so we need to determine the ideal optimal solution based on the principle, which makes unit structure compact and has the fastest response rate. We took a measure when the multiple-changed group get to short focus position, and we chose the $X_{31}$ as compensation curve until the group arrive to -1, and $m_{2.1} = m_{2.2} = -1$. The compensating curve will convert to new curve $X_{32}$.

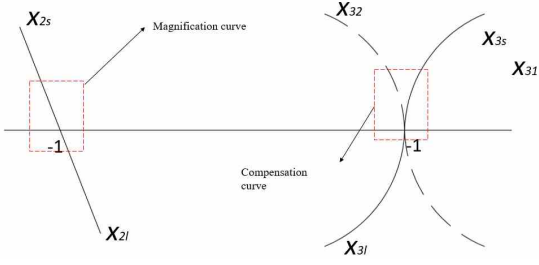

Fig. 2 The movement curve of lens

## B. Probe release

Fig.3 shows the detailed structure of this new probe. A double-sided fully symmetrical design is employed for ensuring that the optical paths on both sides are consistent. It further ensures that the light reaching the surface of the object has exactly the same patterns in terms of spot size, incident angle, and spot distance. Next, we will introduce the core components and their functions, respectively. Component 1 (abbreviated as C1) -- Connection flange, whose functions is to mount the stepper motor and the fiber holder on the flange with screws, and then fixate the flange on the outer casing. C2 -- The motor provides kinetic power input for the device. The fiber is used to deliver the pulsed laser. The C3 – Spot-adjustable Unit (SAU)

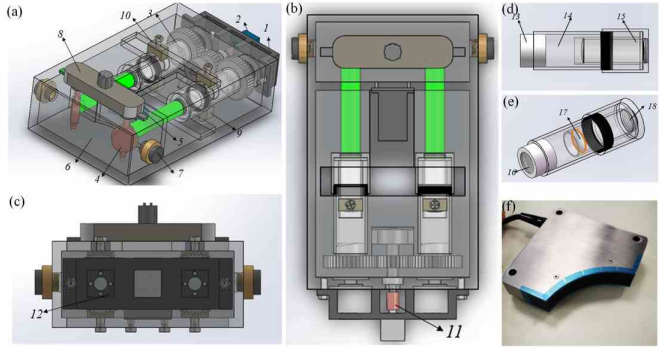

Fig. 3. Probe design. (a)~(c) 3D structure of probe. (d) ~(e)SAU (f) Customized UT(ultrasonic transducer). The detail description of components. (1. Connection flange. 2. stepper motor. 3.SAU 4. Angle-adjustment mirror 5. UT 6.Water tank 7.Water intake/outlet 8. Angle-Adjustment Motor 9. Below bracket 10.upper bracket 11.Coupling 12.Fiber connection 13.Fixed group sleeve 15.Multiple-changed sleeve 16. Compensation sleeve 17. Fixed group mirror 18. Multiple-changed mirror 19. Compensation Mirror)

functions to continuously expand the fiber exiting beams to form pulsed laser beams with different beam diameters. C4 -- Angle adjustment mirror to adjust the angle of the incident light. C5 -- An arc-shaped ultrasonic transducer (UT) that receives PA signals for image reconstruction. C6 -- Water tank. Its function is to provide a propagation medium for the acoustic signal and transparency for light delivery. C7 -- Inlet/Outlet. C8 -- Angle motor for opposite side rotation. Its function is to cooperate with C4 to provide the C4 with a side-to-side opening and closing function. C9, 10 -- SAU upper and lower brackets to fix SAU. C11 -- coupling, connecting the motor and gear shaft for power input. C12 – Fiber connection is a carrier for fiber mounting.

The OPO pulsed laser (OPOTEK, US) is coupled into a customized Y-type fiber, whose light splitting ratio is 1:1. It forms a parallel light with a diameter of 4 mm incident into the SAU[10].

The parallel light is emitted from twofold to sixfold, reflected by the mirrors with the characteristic of opening and closing contralaterally, and then obliquely incident to the surface of the object to be tested. The PA signals are collected by a customized arc-shaped ultrasonic transducer with 60% bandwidth, which integrated 32 channels with center frequency of 2.5MHZ, shown in Fig. 3(f) (Doppler Inc., China). We designed a water tank for the purpose of avoiding severe attenuation of the acoustic signals. The inlet and outlet are designed on both sides of the tank to ensure that the tank is blank when the probe is not working[13].

## III. EXPERIMENTAL TEST

### A. System Setup

To validate the performance of our proposed probe, we built a PA imaging system for experimental testing in our lab[14]. Fig.4 shows the diagram of the whole system. It demonstrates the PA imaging system based on our developed device, which consists of a mechanical structure, control system, data acquisition and processing system[15].The nanosecond pulsed laser is coupled to a customized glass Y-shaped fiber (Zhanrui Inc, China.) that can carry high laser energy with the single multimode core via a fiber coupler. The laser beam from the fiber output is expanded by SAU. Then the beam is reflected by the mirror and subsequently reach the sample surface. The photoacoustic signals detected by the ultrasound scanner were transferred to the 32-channel data acquisition board integrated inside the ultrasound imaging platform for further post-processing and image reconstruction. The trigger out signal of the laser source was used to trigger the ultrasound platform for signal acquisition synchronization[15]. The original PA signals of the sample are collected by the system and processed with the wavelet denoising method bearing adaptive threshold selection system and processed with the wavelet denoising method bearing adaptive threshold selection[16]. and then, DAS algorithm written in Matlab (Mathworks Inc, USA.) is utilized for image reconstruction[14].

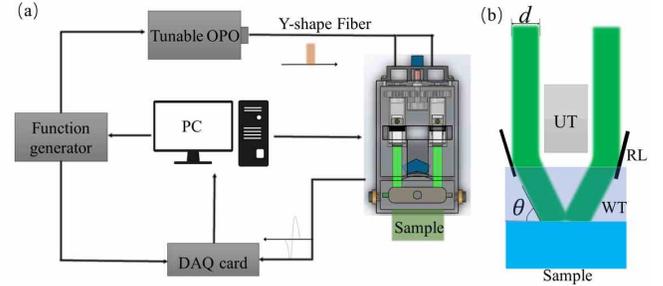

Fig. 4. (a)PA imaging system based on new probe.(b) adjust Parameters. RL: reflect mirror. UT: ultrasonic transducer.

## B. SAU Examination

Continuous laser was used as the source-input to test the function of the SAU. To start with, adjust the laser source, SAU and scale plate so that they were on the same level. Then we used a three-axis precision optical fixture to clamp SAU. Afterwards, we aligned the output laser through two laser-collimate mirrors and adjusted the SAU so that the laser could enter the SAU vertically. Consequently the output beam directly entered the soleplate with the scale. At the end of the light-axis, beam analyzer was used as the experimental recorder to obtain the results of beam size and energy distribution after expansion. The SAU functional verification test device is shown in Fig.5.

## C. Phantom Fabrication

The key capability for imaging under different illumination schemes should be demonstrated. Thus, a vessel phantom was designed for validating our hypothesis. The agar solution with a concentration of about 0.1g/ml was formed, and we slowly poured it into the prepared molding. The agar-abs-agar three-layer structure of the phantom model was then completely manufactured, shown in Fig 6. Moreover, we imaged the phantom via the PA system chosing different schemes. A stepper motor was applied to generate a smoother power input, with the speed set to 1 r/s and a time interval of 5s set every 5r for system imaging. We showed images of combined parameters at different locations and demonstrated the imaging capabilities of the device.

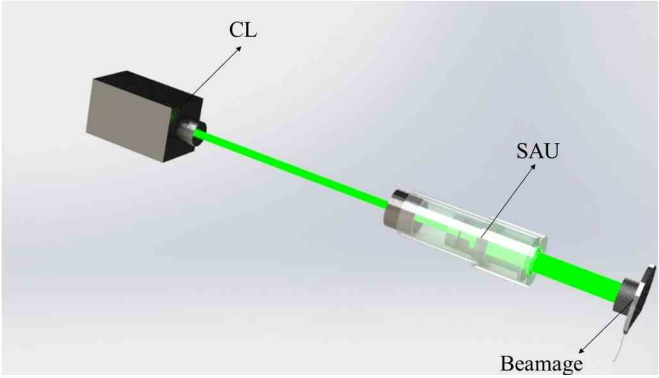

Fig.5 SAU functional verification test

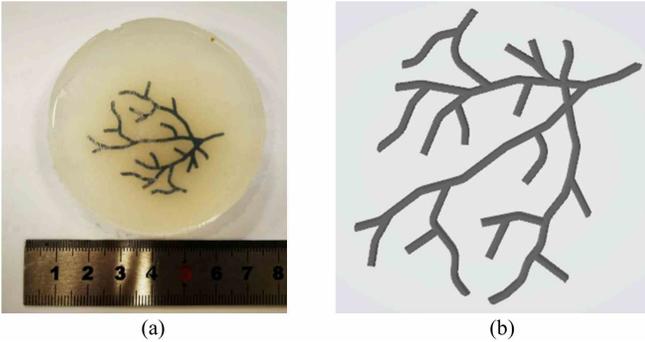

Fig. 6. Phantom image

## IV. RESULTS

### A. SAU Test results

In the beam-expansion experiment, we used a Beam analyzer (BEAMAGE-4M, Spektrum Inc.) to record the various parameters of the output beam spots. Fig. 7(a) and 7(c) show the energy distribution of the shaft with the 3D view after 2x and 3x expansion, and the corresponding detailed parameters of the actual beam are shown in Fig. 7(b) and 7(d). The results show clearly that the incident beam with a diameter of 3 mm is finally expanded to ~6.28 mm and ~9.025 mm, respectively, meeting the predetermined functional requirements.

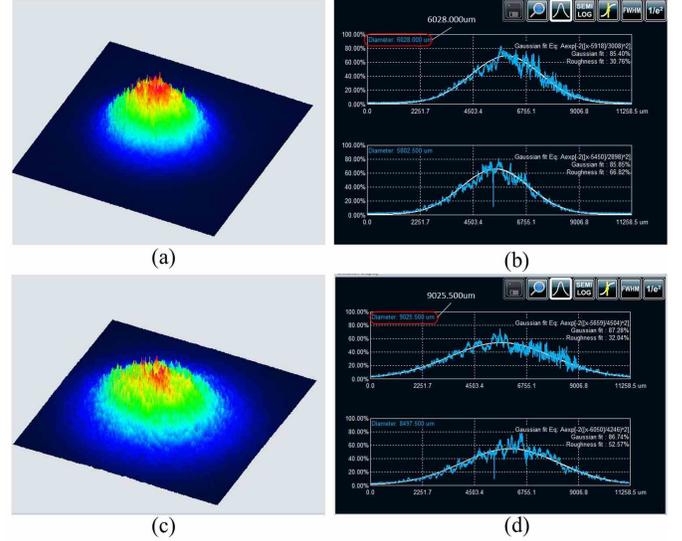

Fig. 7. The SAU test results (a) 2x BA (beam analyzer) shooting picture, (b) 2x spot parameters, (c) 3x BA shooting picture (d) 3x spot parameters.

### B. PA Imaging In Phantom

Based on the phantom we fabricated, we chose the suitable spot whose direction is perpendicular to the surface of the sample. Different illumination schemes could be quickly formed by adjusting the illumination angle and the SAU, hence obtaining different PA images at different angles. The detailed definition of adjusting parameters are given in Fig.4(b), where 'd' is the diameter of beam from SAU, and the '$\theta$' is the incident angel arriving at the sample.

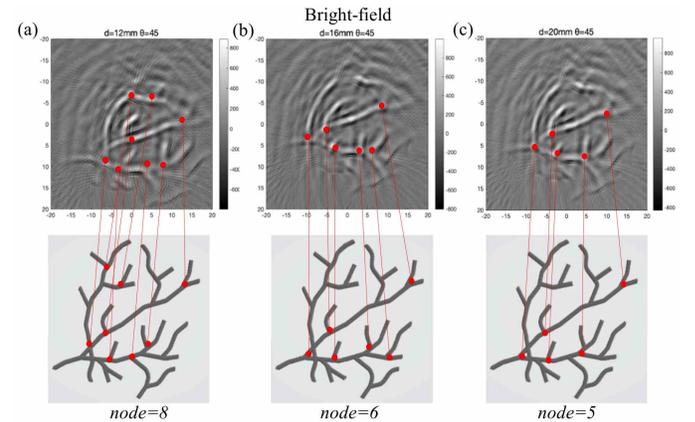

Fig. 8. PA imaging results of phantom. Column(a) d=12mm θ=45. Column(b) d=16mm θ=45. Column(c) d=20mm θ=45.

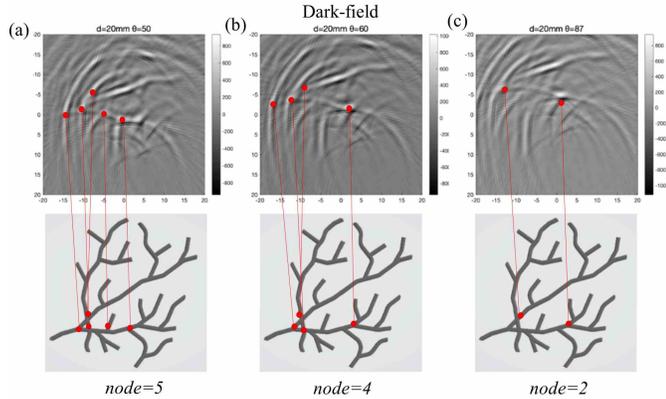

Fig.9 PA imaging results of phantom. Column(a) d=20mm θ=50. Column(b) d=20mm θ=60. Column(c) d=20mm θ=87

Columns (a)~(c) in Fig. 8 and Fig. 9 reveal the PA imaging results in bright-field and dark-field schemes respectively, showing the nodes in PA images and the corresponding real phantom. According to the imaging results, we calculated the contrast in every image, and since it was a phantom of vessels, we also counted the number of nodes, which could tell an important clue called cross node information. Fig.10. demonstrates the contrast in every image. The bar graph explicitly tells that with the incident angle increasing in the dark-field, the poorer contrast and the fewer nodes are obtained. This is because since the phantom was set right under the probe, the bigger central distance between two spots was, the fewer photons would deposit on the phantom. Meanwhile, the highest contrast was obtained when d=12mm and θ=45, when the node number happened to be the biggest as well, up to 8; this could provide more detailed information. We also show the corresponding node graph in Fig.11. Theoretically there are infinite combinations of lighting parameters, which couldn't all be listed. The specific imaging quality is affected by many factors. Users need to combine their own needs and the actual imaging quality to choose the parameter combination.

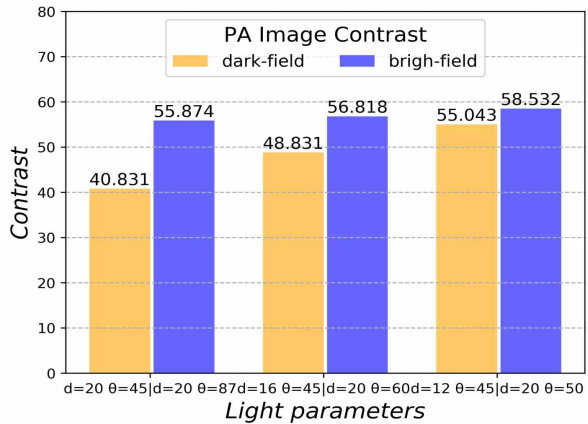

Fig. 10 Contrast of PA images In Fig.8&9

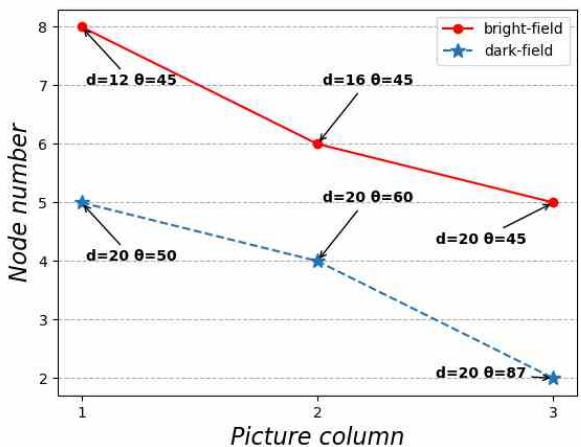

Fig. 11 Corresponding nodes of PA images In Fig.8&9

*C. PA Imaging In-Vivo*

All animal experiments strictly follow the Institutional Animal Care and Use Committee of ShanghaiTech University. The 1 mg/ml ICG (Indocyanine Green) solution was pre-deployed. 5 mg ICG powder was sufficiently dissolved in physiological saline[17]. Choosing Balb/c nude mice (age: 4-week, male, weight: 20~30g ) as experiment animals, we anesthetized them with drug (25g of tribromoethanol and 15.5ml of tert-amyl alcohol were thoroughly mixed and stored as a storage solution in the dark, and at work, 0.5ml of 1.2% storage solution and 3.5ml of 0.9% physiological saline, intraperitoneal injection, 20ug/g weight) under sterile conditions and then fixed them in the detection area with the

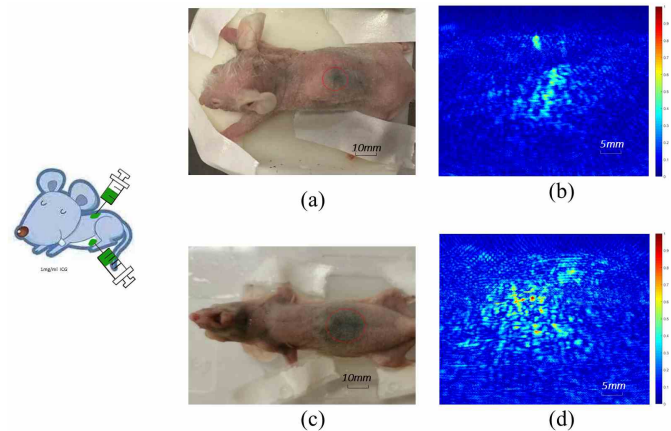

Fig. 12 In vivo PA imaging results of rat: (a)~(c) Back and belly pictures after ICG subcutaneous injection . (b) PA image of rat back . (d) PA image of rat belly

supine and lying posture, as shown in Fig.12 (a) and (c) [18].

We injected the diluted ICG solution into the rat back and belly by subcutaneous injection (inject with uniform speed using a syringe, 0.5mg/s). We controlled the dosage that in the belly is twice the back. The 800-nm wavelength laser, matching the peak optical absorption of ICG, was utilized for photoacoustic signal excitation throughout the experiment. The maximum laser energy on the rat skin was approximately 16 mJ/cm2, far below the ANSI (American National Standards Institute) safety standard (30 mJ/cm2) [19] . Fig.12 (a)~(c) show the back and belly photos after ICG injection. It can be clearly observed that the area with distribution of ICG in the belly is about twice of that in the back. The PA images obtained by the PA system in the bright-field scheme are shown in Fig. 12 (b) and 10(d), whose corresponding illumination scheme was determined after multiply adjusting the incident angle and spot-size. The results suggest that the PA images in back and belly are different, and the distribution of different positions basically matches that of the real rat.

## V. Conclusion And Discussion

In theory, countless excitation schemes can be realized by smoothly changing the diameter of the incident spot and the angle of incidence. For the PAT system, multiple solutions can be implemented in a limited space, which can save economic and time costs. Furthermore, we can find the best scheme of sample-domain after a limited number of changes. We are establishing a closed-loop automatic optimal control system to implement image-to-signal feedback control. Thus, the goal of automatically selecting the optimal image is achieved.

In this paper, a novel light-adjustable portable probe was proposed for scheme-optimized PA imaging. The function is realized that multiple parameters of the exiting beam are enabled to be adjustable. As the incidence angle of illumination changes, the requested light schemes can be obtained accordingly. Moreover, we tested the SAU in the beam expansion capability based on the CCD test platform, showing that the newly designed SAU can perform well. We then built a PAT system with the probe, obtaining PA images of a 3D-print vessel phantom in different illumination schemes, which showed the pros and cons of functioning in different fields. At the end of the paper, we introduced the in vivo experiments injecting the ICG solution into the back and belly of Balb/c rat and imaging the interested areas. The phantom and in vivo imaging results suggest that the proposed probe makes the PA imaging system more compact, diverse, and versatile.

## References

[1] M. Xu and L. V. Wang, "Photoacoustic imaging in biomedicine," *Review of Scientific Instruments,* vol. 77, no. 4, p. 041101, 2006.

[2] L. V. Wang, "Multiscale photoacoustic microscopy and computed tomography," *Nature Photonics,* vol. 3, no. 9, pp. 503-509, 2009/09/01 2009.

[3] F. Xiaohua, G. Fei, and Z. Yuanjin, "Photoacoustic-Based-Close-Loop Temperature Control for Nanoparticle Hyperthermia," *IEEE Transactions on Biomedical Engineering,* vol. 62, no. 7, pp. 1728-1737, 2015.

[4] F. Gao, X. Feng, and Y. Zheng, "Coherent Photoacoustic-Ultrasound Correlation and Imaging," *IEEE Transactions on Biomedical Engineering,* vol. 61, no. 9, pp. 2507-2512, 2014.

[5] T. Duan, H. Lan, H. Zhong, M. Zhou, R. Zhang, and F. Gao, "Hybrid multi-wavelength nonlinear photoacoustic sensing and imaging," *Optics Letters,* vol. 43, no. 22, pp. 5611-5614, 2018/11/15 2018.

[6] Y. Zhao, D. Jiang, H. Lan, and F. Gao, "Image Infusion of Photoacoustic Imaging Based on Novel Adjustable Hand-held Probe," in *2019 IEEE International Ultrasonics Symposium (IUS)*, 2019, pp. 2366-2368.

[7] M. Li *et al.*, "Linear array-based real-time photoacoustic imaging system with a compact coaxial excitation handheld probe for noninvasive sentinel lymph node mapping," *Biomedical Optics Express,* vol. 9, no. 4, pp. 1408-1422, 2018/04/01 2018.

[8] G. S. Sangha, N. J. Hale, and C. J. Goergen, "Adjustable photoacoustic tomography probe improves light delivery and image quality," *Photoacoustics,* vol. 12, pp. 6-13, 2018/12/01/ 2018.

[9] Y. Bai, B. Cong, X. Gong, L. Song, and C. Liu, "Compact and low-cost handheld quasibright-field linear-array probe design in photoacoustic computed tomography," *Journal of Biomedical Optics,* vol. 23, no. 12, p. 121606, 2018.

[10] F. Gao, X. Feng, and Y. Zheng, "Advanced photoacoustic and thermoacoustic sensing and imaging beyond pulsed absorption contrast," *Journal of Optics,* vol. 18, no. 7, p. 074006, 2016/05/31 2016.

[11] M. W. Schellenberg and H. K. Hunt, "Hand-held optoacoustic imaging: A review," *Photoacoustics,* vol. 11, pp. 14-27, 2018/09/01/ 2018.

[12] C. Tao, "Zoom optical system design, National Defense Industry Press（Chinese）," 1988.

[13] Y. Zhao, S. Yu, B. Tao, and F. Gao, "Adjustable Handheld Probe Design for Photoacoustic Imaging:Experimental Validation," in *2019 41st Annual International Conference of the IEEE Engineering in Medicine and Biology Society (EMBC)*, 2019, pp. 7119-7122.

[14] H. Lan, T. Duan, H. Zhong, M. Zhou, and F. Gao, "Photoacoustic Classification of Tumor Model Morphology Based on Support Vector Machine: A Simulation and Phantom Study," *IEEE Journal of Selected Topics in Quantum Electronics,* vol. 25, no. 1, pp. 1-9, 2019.

[15] D. Jiang, H. Lan, H. Zhong, Y. Zhao, H. Li, and F. Gao, "Low-Cost Photoacoustic Tomography System Based on Multi-Channel Delay-Line Module," *IEEE Transactions on Circuits and Systems II: Express Briefs,* vol. 66, no. 5, pp. 778-782, 2019.

[16] M. Zhou, H. Xia, H. Lan, T. Duan, H. Zhong, and F. Gao, "Wavelet de-noising method with adaptive threshold selection for photoacoustic tomography," in *2018 40th Annual International Conference of the IEEE Engineering in Medicine and Biology Society (EMBC)*, 2018, pp. 4796-4799.

[17] M. Capozza *et al.*, "Photoacoustic imaging of integrin-overexpressing tumors using a novel ICG-based contrast agent in mice," *Photoacoustics,* vol. 11, pp. 36-45, 2018/09/01/ 2018.

[18] W. Liu and J. Yao, "Photoacoustic microscopy: principles and biomedical applications," (in eng), *Biomedical engineering letters,* vol. 8, no. 2, pp. 203-213, 2018.

[19] R. J. Thomas, B. A. Rockwell, W. J. Marshall, R. C. Aldrich, S. A. Zimmerman, and R. J. Rockwell, "A procedure for multiple-pulse maximum permissible exposure determination under the Z136.1-2000 American National Standard for Safe Use of Lasers," *Journal of Laser Applications,* vol. 13, no. 4, pp. 134-140, 2001.